\documentclass[10pt,conference]{IEEEtran}
\IEEEoverridecommandlockouts
\usepackage{cite}
\usepackage{amsmath,amssymb,amsfonts}
\usepackage{algorithmic}
\usepackage{graphicx}
\usepackage{textcomp}
\usepackage{xcolor}
\usepackage{booktabs} 
\usepackage{listings}
\usepackage{multirow}
\usepackage{soul}
\usepackage{xspace}
\usepackage{graphicx}
\usepackage{ifthen}
\usepackage[normalem]{ulem} 
\usepackage{xcolor,colortbl}
\usepackage{hyperref}

\newboolean{showedits}
\setboolean{showedits}{true} 
\ifthenelse{\boolean{showedits}}
{
	\newcommand{\del}[1]{\textcolor{red}{\sout{#1}}} 
	\newcommand{\nbe}[3]{
		{\colorbox{#3}{\bfseries\sffamily\scriptsize\textcolor{white}{#1}}}
		{\textcolor{#3}{\sf\small$\blacktriangleright$\textit{#2}$\blacktriangleleft$}}}
}{
	\newcommand{\del}[1]{} 
	
	\newcommand{\nbe}[3]{}
}

\newboolean{showcomments}
\setboolean{showcomments}{true} 
\newcommand{\id}[1]{$-$Id: scgPaper.tex 32478 2010-04-29 09:11:32Z oscar $-$}

\ifthenelse{\boolean{showcomments}}
 {
 	\newcommand{\nbc}[3]{
 		{\colorbox{#3}{\bfseries\sffamily\scriptsize\textcolor{white}{#1}}}
		{\textcolor{#3}{\sf\small$\blacktriangleright$\textit{#2}$\blacktriangleleft$}}}
	
 }{
 	\newcommand{\nbc}[3]{}
 	
 }

\usepackage[most]{tcolorbox}
\ifthenelse{\boolean{showedits}}
{
  \newtcolorbox{inserted}{%
       title=Inserted text:,
       colframe=blue,colback=blue!5!white,
       breakable,
       leftrule=0mm, 
       bottomrule=0mm,
       rightrule=0mm,
       toprule=0mm,
       arc=0mm, outer arc=0mm,
       oversize
  }
  \newtcolorbox{deleted}{%
       title=Deleted text:,
       colframe=red,colback=red!5!white,
       breakable,
       leftrule=0mm, 
       bottomrule=0mm,
       rightrule=0mm,
       toprule=0mm,
       arc=0mm, outer arc=0mm,
       oversize
  }
  \newtcolorbox{refactored}{%
       title=Rewritten text:,
       colframe=blue,colback=red!5!white,
       breakable,
       leftrule=0mm, 
       bottomrule=0mm,
       rightrule=0mm,
       toprule=0mm,
       arc=0mm, outer arc=0mm,
       oversize
  }
}{

}
\newboolean{isblinded}
\setboolean{isblinded}{true}
\ifthenelse{\boolean{isblinded}}
{\newcommand\blind[1]{BLINDED\xspace}}
{\newcommand\blind[1]{#1\xspace}}

\newcommand{\commented}[1]{}

\newcommand{\eg}{\emph{e.g.,}\xspace}
\newcommand{\ie}{\emph{i.e.,}\xspace}
\newcommand{\etal}{\emph{et al.}\xspace}

\definecolor{source}{gray}{0.95}
\usepackage[T1]{fontenc} 

\newcommand{\boxit}[1]{\vspace{0.2cm}
\noindent
\fbox{
\begin{minipage}{8.3cm}
\footnotesize \emph{#1} 
\end{minipage}
}
}

\lstdefinelanguage{Java}{
  tabsize=4
}[keywords,comments,strings]

\definecolor{source}{gray}{0.95}
\definecolor{highlight}{gray}{0.9}

\lstnewenvironment{codesnippet}{%
	\lstset{%
		frame=single,
		framerule=0pt,
		mathescape=false
	}
}{}

\newcommand{\GH}{GitHub\xspace}
\newcommand{\SO}{Stack Overflow\xspace}

\begin{document}
\title{Worrisome Patterns in Developers: A Survey in Cryptography}
%
%


\author{\IEEEauthorblockN{Mohammadreza Hazhirpasand}
\IEEEauthorblockN{Oscar Nierstrasz}
\IEEEauthorblockA{University of Bern\\
Bern, Switzerland}
\and
\IEEEauthorblockN{Mohammad Ghafari}
\IEEEauthorblockA{
University of Auckland\\
Auckland, New Zealand\\
m.ghafari@auckland.ac.nz
}
}

%
\maketitle              
\begin{abstract}
We surveyed 97 developers who had used cryptography in open-source projects, in the hope of identifying developer security and cryptography practices.
We asked them about individual and company-level practices, and divided respondents into three groups (\ie high, medium, and low) based on their level of knowledge.
%
%
We found differences between the high-profile developers and the other two groups.
For instance, high-profile developers have more years of experience in programming, have attended more security and cryptography courses, have more background in security, are highly concerned about security, and tend to use security tools more than the other two groups.
Nevertheless,
we observed worrisome patterns among all participants such as the high usage of unreliable sources like \SO, and the low rate of security tool usage.
\end{abstract}

\section{Introduction}
\label{sec:intro}
Many developers do not correctly use cryptographic (crypto) APIs~\cite{hazhirpasand2019impact}.
Investigation of developer questions on the Stack Overflow website showed that they often struggle with understanding cryptography concepts \cite{hazhirpasand2021hurdles}.
%
Researchers have developed new tools and APIs to ease the adoption of cryptography \cite{kafader2021}, yet crypto issues are prevalent~\cite{hazhirpasand2020java}.

We aim to shed light onto the following research question: \emph{``What are the practices of developers who use cryptography in the wild?''}
We conducted an exploratory study to identify developer cryptography and security practices.
We designed an online survey containing questions regarding developer demographics, and developer- and company-level security background.
We sent the survey to 1231 developers who had frequently used crypto APIs in open-source projects.

We received 97 responses.
Most of the respondents (\ie 73\%) had studied Computer Science and all participants claimed to have knowledge in cryptography.
In particular, 60\% stated that they are knowledgeable, and 19\% very knowledgeable.
We divided the respondents into three groups, namely \emph{low, me\-dium, and high-profile developers in cryptography}.
We classified them based on their responses to self-reported knowledge in cryptography (in short, knowledge).
In order to help the participants to choose the most suitable level of knowledge, we provided them with definitions of what each level of knowledge means and then compared the responses among the groups.

We noticed that for developer-level and company-level security factors, \eg security course attendance or existence of security consultant at work, the high-profile developers have considerably better records than the other groups.
In the same vein, developers with medium and low profiles have similar responses.
With respect to company-level factors, we observed low attention towards security-related factors, \eg half of the companies have no security training or only every two years, and 71\% of companies have no security consultant.
We therefore arrive at the conclusion that the high-profile group tends to report stronger security and crypto-related characteristics than those who did not feel confident in their knowledge.
Even though there exists a high rate of security concerns among developers and companies, there are alarming findings, \eg low acceptance rate of security tool adoption or relying on unreliable sources, \eg \SO, to solve crypto problems.
Therefore, practitioners and researchers should consider the potential effects and devise proper guidelines, training, and methods to lead inexperienced developers in this domain.
Future work should assess developer knowledge based on their real performance in a controlled experiment, the severity of security policies enforced in companies, and project-level security requirements.

\section{Developer Survey}
\label{sec:devsurvey}

We conducted an online survey with developers identified in a recent work ~\cite{hazhirpasand2020java} as having used Java Crypto APIs in real-world applications,  to understand what security and cryptography practices such developers report. In the following subsections, we explain our methodology and present the survey results.



\subsection{Methodology}
\label{sub:surveya}
We adopted an anonymous online survey approach which involves the following steps: (i) selecting developers, (ii) designing the survey, (iii) testing and publishing the survey, and (iv) analyzing the survey.
The questions and the responses of the survey are available online.\footnote{\url{http://crypto-explorer.com/crypto/}}

\subsubsection{Objective}
\label{sub:obj}
Previous studies have shown that developers have difficulty in using cryptography securely~\cite{hazhirpasand2019impact}\cite{nadi2016jumping}.
In this study, we pose the following research question: \emph{``What are the practices of developers who use cryptography in the wild?''} to tackle the reasons why developer performance varies in using cryptography.
The objectives of this research are as follows:
(1) except for technical difficulties of crypto APIs explored in previous research~\cite{Egele}~\cite{hazhirpasand2020java} \cite{muslukhov2018source}, the findings can shed some light on the worrisome and promising practices among developers with respect to cryptography,
(2) finding the indicative factors can assist professionals to correctly guide and lead such developers at workplaces.

\subsubsection{Selecting developers}
We selected developers from a recent study conducted by Hazhirpasand \etal~\cite{hazhirpasand2020java}, where the authors investigated Java Cryptography Architecture (JCA) uses and misuses in 489 open-source projects.
They identified developers who committed code containing crypto uses to these repositories, \ie they extracted their names and email addresses using the git blame command.

%



\begin{table}[]
\footnotesize	
\centering
\caption {Factors to explore in the survey} \label{tab:factors} 
\begin{tabular}{|l|l|l|}
\hline
\textbf{Demographics}                                                        & \textbf{Developer-level}                                                                        & \textbf{Company-level}                                                             \\ \hline
Developer age                                                                & \begin{tabular}[c]{@{}l@{}}Security course \\ attendance and experience\\ as code auditor\end{tabular}   & \begin{tabular}[c]{@{}l@{}}Security training\\ by company\end{tabular}                      \\ \hline
\begin{tabular}[c]{@{}l@{}}Years of experience\\ in programming\end{tabular} & \begin{tabular}[c]{@{}l@{}}Crypto knowledge level\\ and experience in using \\ Crypto API\end{tabular}   & \begin{tabular}[c]{@{}l@{}}Existence of security\\ consultant\end{tabular}                  \\ \hline
\begin{tabular}[c]{@{}l@{}}Years of experience\\ in Java\end{tabular}        & \begin{tabular}[c]{@{}l@{}}Background in IT security\\ and security concern level\end{tabular}           & \begin{tabular}[c]{@{}l@{}}Company security \\ concern level\end{tabular}                   \\ \hline
Educational level                                                            & \begin{tabular}[c]{@{}l@{}}Ways of solving crypto \\ problems and evaluating \\ crypto code\end{tabular} & \begin{tabular}[c]{@{}l@{}}The percentage of \\ security developer \\ in company\end{tabular} \\ \hline
\end{tabular}
\end{table}

\subsubsection{Survey Design}
In the survey, we collected information about the participants in three sections, and then evaluated the factors to determine which of them influence developer knowledge.
To determine the explored factors, we studied the literature and identified studies wherein individual or work-related facets were studied with regard to cryptography (See \autoref{tab:factors}). 
Thereafter, we constructed a list of explored factors that could influence developer knowledge, namely security tool adoption \cite{johnson2013don} ~\cite{ayewah2008report}, security concern \cite{xiao2014social} \cite{thomas2018security}, means of resolving crypto challenges \cite{acar2016you}  \cite{poller2017can}, security training and its frequency \cite{valentine2006enhancing} \cite{nadeem2014computer}  \cite{gardner2014building}, and work and technical experience \cite{robillard2009makes} \cite{acar2017security} \cite{oliveira2018api} \cite{acar2016you}. 
Nevertheless, the aim of the previous studies was to evaluate developer performance or developer practice but not developer knowledge. 

The initial section is dedicated to the demographic information of developers.
Within this section, we asked for their degree, field, age, years of programming and Java programming experience.
Participant demographics help us to determine which factors may affect a respondent's answers.

In the second section, we mostly focus on developer practices, \eg security or cryptography course attendance.
They are asked to specify their level of knowledge about cryptography as well as their experience with crypto APIs.
We used a 5-point Likert scale to ask developer security concerns in development. Further, 
We inquired developers what information sources they use to solve a crypto scenario or how they evaluate a crypto code snippet.

In the last section of the survey, we primarily concentrate on company-level factors. 
We provided them with questions regarding the existence of security consultant, company-level security concern (5-point Likert scale), and the percentage of developers responsible for
secure development. 

\subsubsection{Testing, and Publishing the Survey Tool}
We used Google Forms to create our online questionnaire.
As overlong questionnaires are commonly not completed on the internet \cite{batinic200011}, we limited the completion time of the survey to less than 5 minutes.
To evaluate the survey before asking the real participants,
we asked five colleagues to review the survey to reveal potential misunderstandings.
Then, based on the received recommendations, we refined the questions and rearranged them.
Next, we emailed the 1231 developers.
We noticed that 128 email addresses were not valid, which left us with 1\,103 potential survey participants.

\subsubsection{Survey Analysis}

We received 97 responses (8.7\%) within a month.
To perform the analysis, we do not consider missing values in the analysis.
We use percentage graphs in order to analyze responses of Likert scale questions.
Notably, the explanations of respondents to the open-ended question consisted of fewer than 20 words. 
However, to minimize human errors, two authors of the paper coded the responses and cross-checked the consistency of the results.

\subsubsection{Knowledge factor}
Nadi \etal conducted two surveys with 48 developers and devised a four-level classification for developer crypto knowledge \cite{nadi2016jumping}.
We used the same levels in the survey.
However, we attempted to minimize the impact of wrong assumptions with regard to how developers report their level of knowledge in cryptography.
As a result, we provided the participants with an explanation of what each level means in this study.
The four-level items can be viewed in the survey file.\footnote{\url{http://crypto-explorer.com/crypto/}}

\subsubsection{Ethics}
The developers' email addresses were identified by the use of the git blame command in a recent study~\cite{hazhirpasand2020java}.
We also asked developers to read the statement of the survey and state their agreement before participating. 
Moreover, we did not collect any personal information except for the information explicitly gathered by the survey instrument.


\section{Developer practice}
\label{sec:discussion}
We discuss our findings from the developer self-reported knowledge perspective.
We then compare variables related to each participant characteristic with the reported knowledge level.
The participants rated their knowledge in cryptography as 21\% (\ie 21) \textit{somewhat knowledgeable}, 60\% (\ie 58) \textit{knowledgeable}, and 19\% (\ie 18) \textit{very knowledgeable}. 
We respectively assign these participants to low (21), medium (58), and high-profile (18) groups in this domain.

\subsection{Developer demographics}
\begin{figure}[h]
\centering
\includegraphics[width=1\linewidth,trim=4 4 4 4,clip]{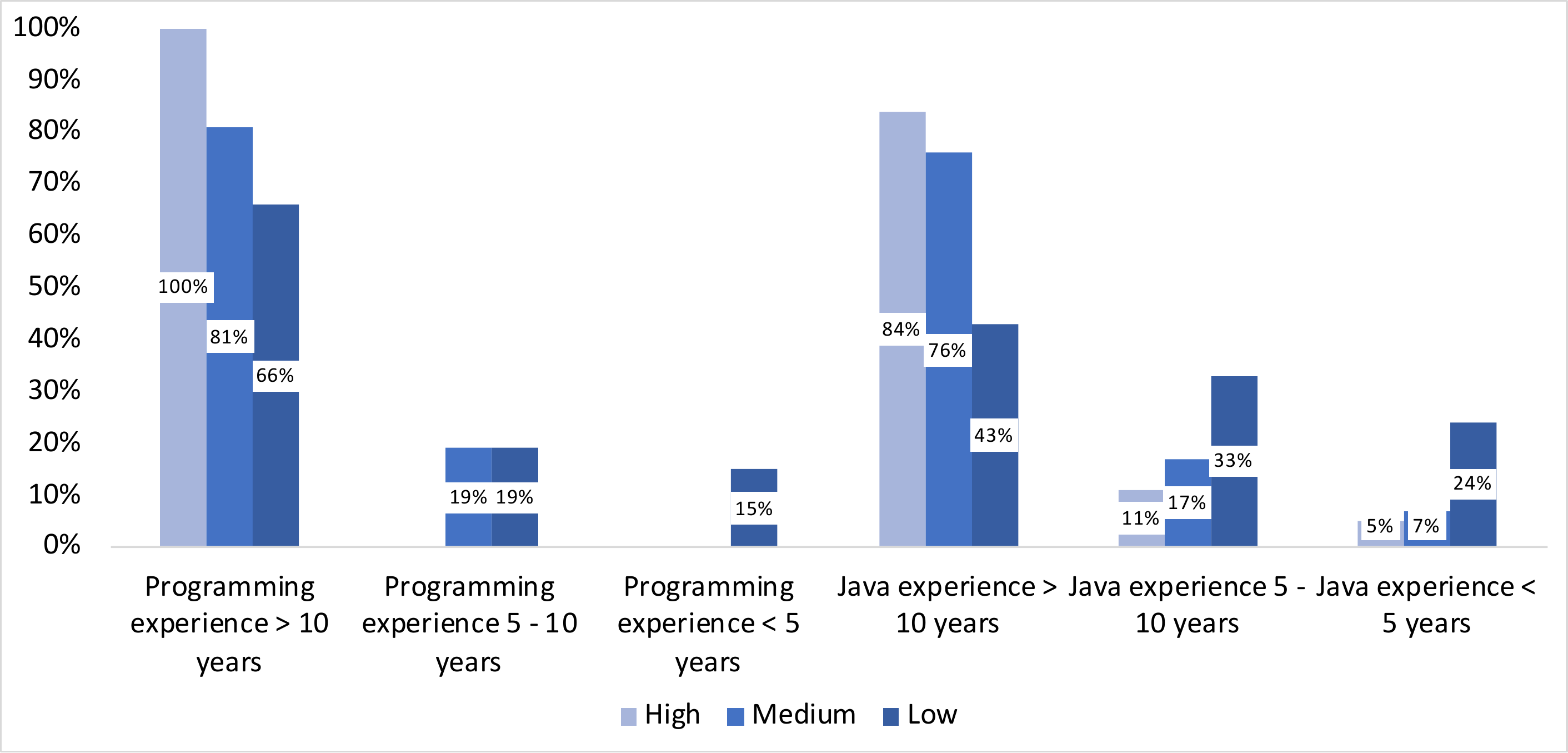}
\caption{Years of experience in programming and Java} 
\label{fig:yrsexp}
\end{figure}

We analyzed whether age, experience in programming, or education are correlated to knowledge in cryptography.
Unsurprisingly, the older participants are, the more experienced they are in programming.
Although experienced participants exist in every knowledge group,
there is a clear distinction between high to medium-profile developers and low-profile developers (See \autoref{fig:yrsexp}).
All participants from the high-profile group have more than 10 years of experience in programming, and there are no participants from high or medium-profile groups with fewer than 5 years of experience in programming. 
The same pattern was seen among the groups for years of programming experience in Java.
The education level is almost evenly distributed among the three groups.
Of the seven Ph.D. participants, five belong to the medium-profile group.

\boxit{
Developers with a high or medium level of knowledge in cryptography have more years of experience in programming and Java.}

\subsection{Developer characteristic}
To have a glance at developer characteristic, we examined the relationship between developer knowledge and any of the following: security course attendance, experience in using crypto libraries, background in IT-security, developer security concern, ways of solving crypto problems and evaluating a crypto code, and working as a source code auditor.

With regards to security or cryptography course attendance,
the more knowledge developers had, the more courses they attended.
The high group has the highest number of participants (22\%) attending \textit{both courses}, while there are 9\% and 14\% such participants of the medium and low groups, respectively.
The number of participants who attended such courses is below 40\% in all groups.


Although several participants (\ie 15) responded that they worked as a secure code auditor, they belong to different knowledge groups, \ie high (3), medium (10), and low (2). 
Just over a half (\ie 53\%) of such participants never attended a security or cryptography course, whereas the rest (\ie 7) attended security and cryptography courses.

In total, 38\% of participants (\ie 37) reported that they had background in IT security.
We observed that a very large proportion (86\% and 61\%) of the low and medium-profile participants had no background in IT security. 
In contrast,  61\% of the high-profile developers expressed their security-relevant background, achieved through various methods \eg bachelor and master thesis on software security, or personal enthusiasm and self-study in software security. 

Notably, all respondents from the high-profile group stated that they have at least two years of experience in using crypto libraries, whereas
 52\% and 21\% of the low and medium-profile developers are not highly experienced (\ie \textless= 2 years) with such crypto libraries.

We received 47 responses regarding the hindrances developers encounter when dealing with cryptographic tasks.
The majority of them mentioned two key obstacles:
the first was the high complexity of using crypto APIs.
For instance, a developer mentioned that \textit{``Wide variety of configuration options''} is troublesome.
The second obstacle concerned unreliable sources and lack of security experts in teams.
For example, one participant blamed \textit{``Poor Java docs''}.
More than half of the 47 respondents (63\%) were from medium-profile developers, and 17\% of them were from the high-profile group.
This means that developers who still feel confident about their knowledge in cryptography struggle to use them in the wild.

\boxit{A large proportion of low-profile developers have no IT security background, while more than half of the high-profile developers do have such backgrounds.
High-profile developers have slightly better records in the security/cryptography course attendance and considerably more years of experience in using crypto libraries than other developers, whereas medium and low-profile developers are almost similar.
developers from all groups mainly complained about the complexity of crypto APIs and insufficient documentation.}

\begin{figure}[h]
\centering
\includegraphics[width=0.9\linewidth,trim=4 4 4 4,clip]{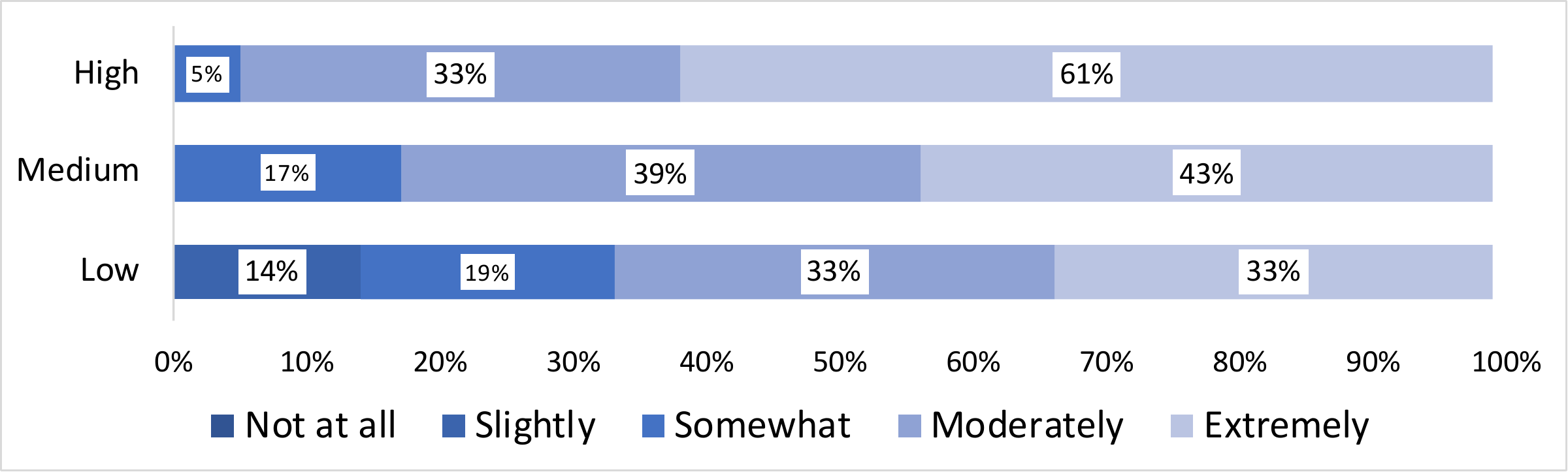}
\caption{Security concern by participants}
\label{fig:devconcern}
\end{figure}

With regards to developer security concerns, 81\% of developers rated their concern as \emph{important} or \emph{very important}.
Remarkably, 61\% high-profile developers are \emph{very} concerned with security while 33\% and 43\% from low and medium groups reported the same level of concern (See \autoref{fig:devconcern}).
Only one high-profile developer is \emph{somewhat} concerned with security while 19\% of low- and 17\% of medium-profile developers reported the same level.

Participants reported various information sources to solve the challenges in a cryptography-related task (presented in \autoref{tab:resources}).
In particular, the primary information sources seem to be online.
In each of the three groups, \textit{websites found on search engines} and \textit{Stack Overflow} are among the top three preferred approaches.
It is noteworthy that the role of information security experts is not as commonly cited as other approaches.
Participants only from high- and medium-profile groups consult with a security expert, and none of them are from the low-profile group.
High-profile participants prefer discussion with colleagues, security consultants, and crypto stack exchange more than other participants.

\begin{figure}[h]
\centering
\includegraphics[width=0.9\linewidth,trim=4 4 4 4,clip]{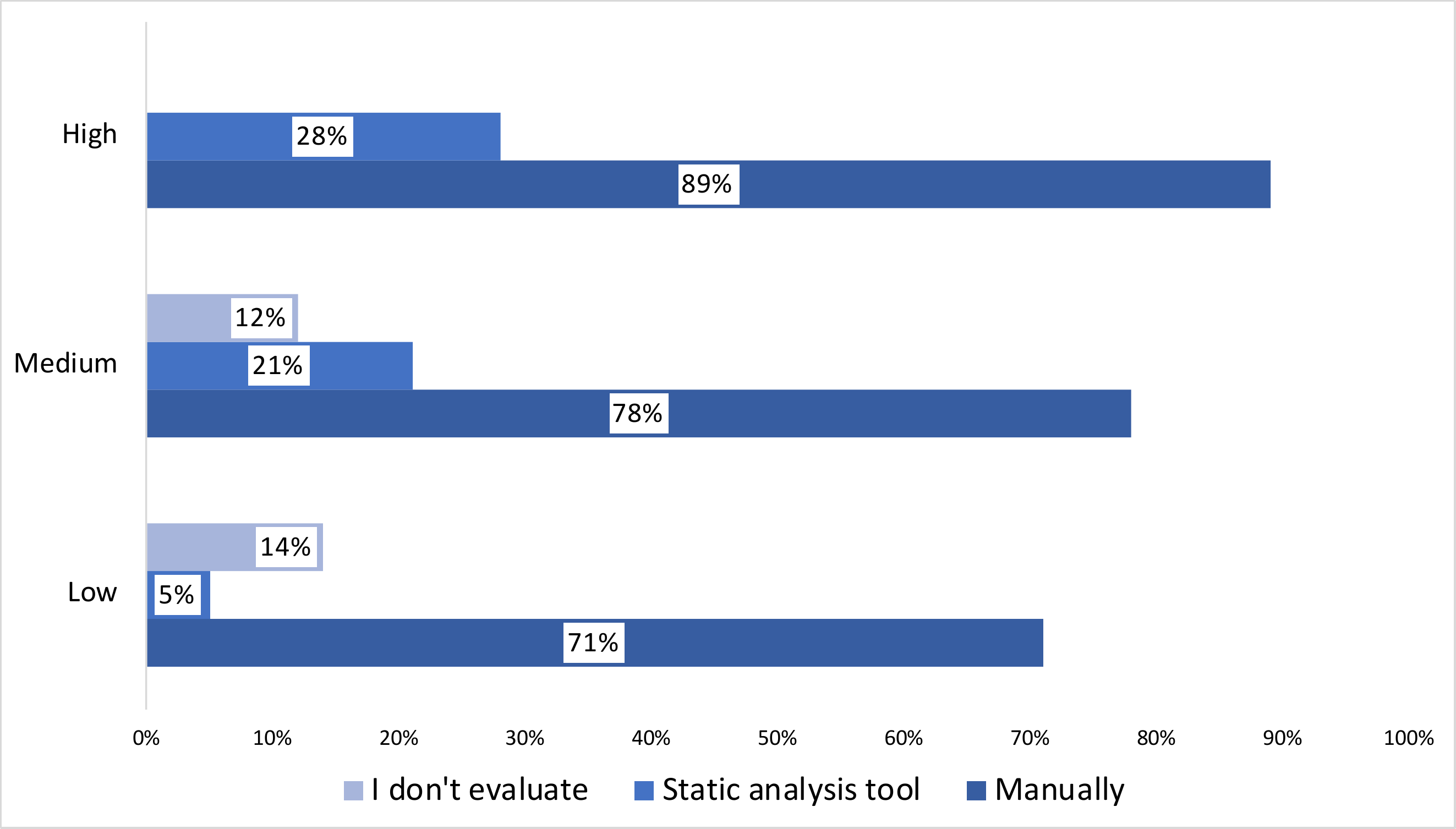}
\caption{How developers evaluate a crypto copy-pasted code}
\label{fig:codeev}
\end{figure}

Just over half of the developers in all groups \textit{only} evaluate their code manually (See \autoref{fig:codeev}).
Remarkably, the total number of participants who use a static analysis tool is fewer than one-fifth (\ie 18) of the total number of participants.
In more detail, high-profile developers had the highest usage of static analysis tools (28\%), while low-profile developers had the lowest usage of such tools (5\%).
Of ten developers who do not evaluate, only one belongs to high-profile group.

\boxit{
Unlike others, nearly all (\ie 17) high-profile developers are \emph{extremely} or \emph{moderately} concerned about security.
Developers mainly solve their crypto problems on Stack Overflow or websites returned by search engines.
The high-profile group benefits more from security consultants, discussions with colleagues, and crypto Stack Exchange to resolve crypto problems.
Developers mainly evaluate crypto-related issues manually rather than using analysis tools.
All high-profile developers evaluate their crypto code and they use static analysis more than others.
In contrast, low-profile developers tend to use static analysis tools less than others, and 16\% do not evaluate their code.
}

\begin{table}[h]
\footnotesize	
\centering
\caption {The information sources that developers use - bold items are the highest} \label{tab:resources} 
\begin{tabular}{llll}
\hline
\textbf{}                  & \textbf{High   } & \textbf{Medium  } & \textbf{Low  }  \\ \hline
Websites on search engines & 83\%          & 88\%            & \textbf{86\%} \\
Stack Overflow             & 72\%          & \%74            & \textbf{\%80} \\
Crypto Stack Exchange      & \textbf{34\%} & 33\%            & 19\%          \\
Security consultant        & \textbf{28\%} & 10\%            & 0\%           \\
Books                      & 33\%          & \textbf{41\%}   & 14\%          \\
Discussion with colleagues & \textbf{77\%} & 38\%            & 57\%          \\ \hline
\end{tabular}
\end{table}


%


\subsection{Company characteristic}
\label{sec:cpch}
\begin{figure}[h]
\centering
\includegraphics[width=0.9\linewidth,trim=4 4 4 4,clip]{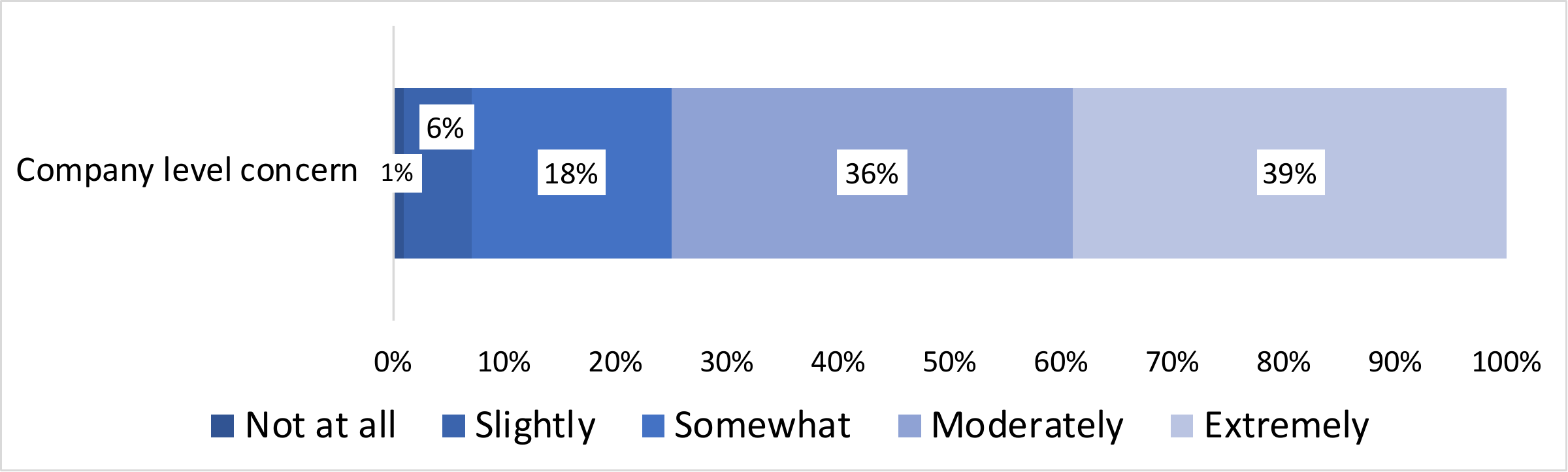}
\caption{Security concerns by companies}
\label{fig:secconcernc}
\end{figure}

Developers were asked to respond to how concerned their companies are with security (See \autoref{fig:secconcernc}).
More than half of the companies (\ie 72) were concerned (important and very important) with security.
Furthermore, 71\% (\ie 69) of the companies do not have any security consultant in their team and 70\% of the companies have no or fewer than 30\% of developers responsible for secure development.
Disappointingly, we learned that 57\% (\ie 53) of respondents do not receive security training at workplace. 
A yearly training interval is the most common approach (27\%), and a two-year training interval is the least common approach (6\%) among companies.

\begin{figure}[h]
\centering
\includegraphics[width=0.8\linewidth,trim=4 4 4 4,clip]{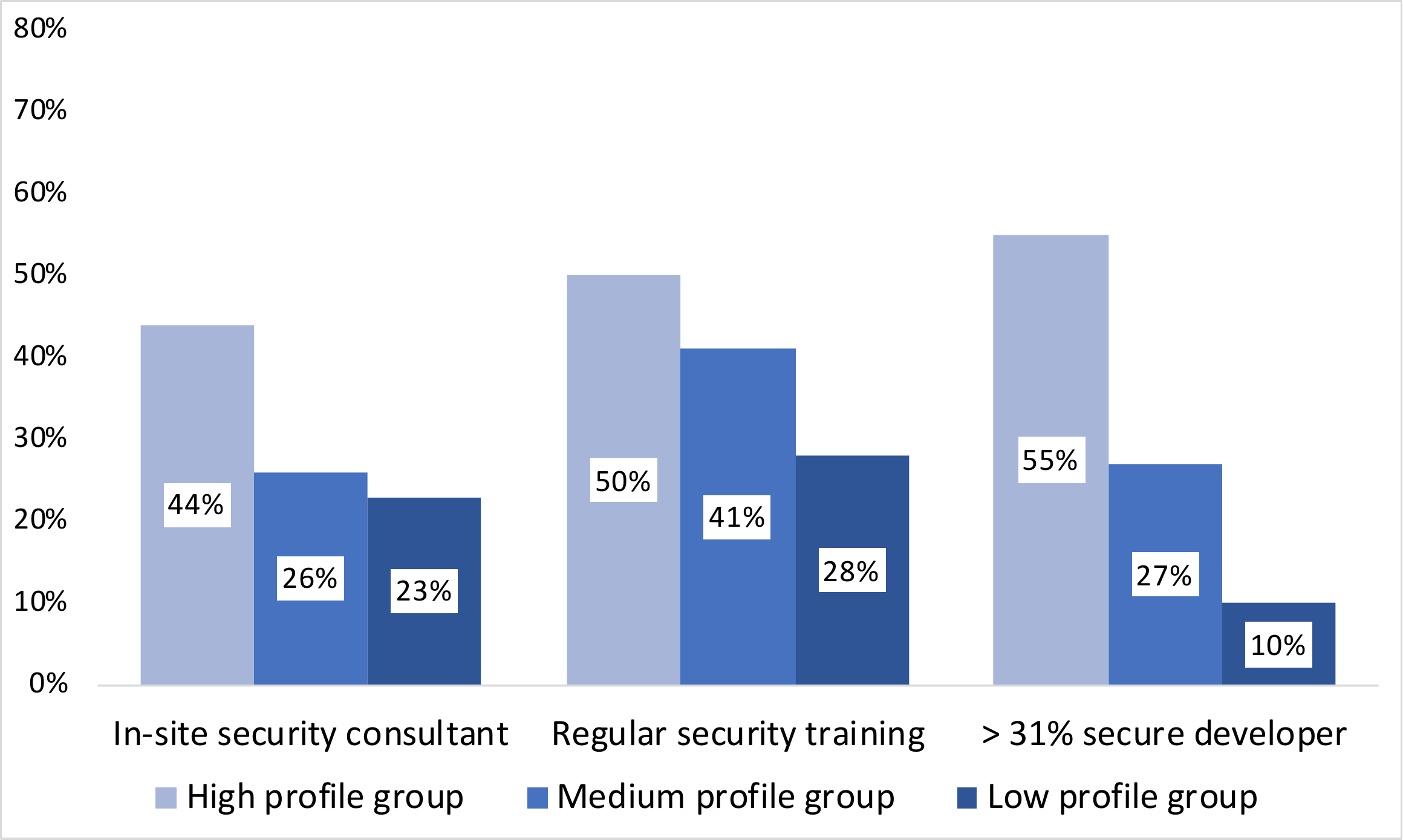}
\caption{Security support provided by participant companies}
\label{fig:seventhree-info}
\end{figure}

We observed the mapping of ``in-site consultant'', ``regular training'', and ``responsible developers'' with developer knowledge (See \autoref{fig:seventhree-info}).
As expected, in all three factors, high-profile developers reported positive responses slightly more than the medium-profile group, and noticeably more than the low-profile group.
However, the medium and low-profile developers are very similar concerning the in-site security consultant.

\boxit{
The majority of companies are concerned with security but the lack of security consultants, regular security training, and security developers are inevitable.
In particular, high-profile developers benefit more from the three factors factors.
}

%

\section{Discussion and Related work}
\label{sec:related}
\emph{\textbf{Developer background:}} 
Acar \etal assigned 307 active GitHub users  to complete several security-relevant programming tasks and surprisingly, found no statistically significant differences concerning functional correctness and security perception among the participants who registered their status as a student, professional developer, or those who had security background~\cite{acar2017security}.
Interestingly, years of experience was not an effective factor for security perception.
Oliveira \etal designed a study for 109 developers to use some APIs that had some blind spots, \ie containing underlying causes to misuse an API, and some easy to use ones \cite{oliveira2018api}.
The results show that developer expertise and experience did not predict their ability to identify blind spots.
In another study, the outcome of an experiment with 54 professional and inexperienced developers for writing security-related code explains that development experience is not a decisive factor for code security \cite{acar2016you}.
Nadi \etal conducted two surveys and asked developers about the issues they face when working with crypto tasks \cite{nadi2016jumping}. 
The participants mentioned several types of issues including lack of documentation, difficulty in API use, and indirection between the APIs and the underlying implementation.
The authors also realized that developers from various knowledge level groups still face the same types of issues in cryptography.
Robillard \etal conducted surveys and interviews with Microsoft developers, and realized that poor documentation is a major learning obstacle for learning APIs \cite{robillard2011field}.
To alleviate unsafe coding practices, security training courses, \eg secure programming, are more effective compared to general security training~\cite{nadeem2014computer}.

\emph{\textbf{Security tool adoption:}} Johnson \etal conducted interviews with 20 developers to understand the determinant factors why static analysis tools were not adopted by many developers \cite{johnson2013don}.
Participants mentioned reasons such as the high rate of false positives, the way that warnings are displayed, faulty integration of the tool into the development process, lack of detailed explanation of bugs with automatic fixes, and not including understandable configuration options in the tool for all levels of developers.
Other researchers investigated the reasons for a low rate of security tool adoption ~\cite{ayewah2008report} \cite{ayewah2008using}.
They found that organization and team policies affect the usage of security-related tools and larger organizations use security tools more than small ones.
The greater adoption of security tools can be influenced by factors such as the culture of the company, security concerns, training, and dedicated security and testing teams.

\emph{\textbf{Ways of solving crypto problems:}} 
Even though using Stack Overflow might help the functional correctness, it leads to more insecure copy-pasted code snippets \cite{acar2016you}.
Ye \etal worked on a system called insecure code snippet detection (ICSD) to detect the imminent insecure code snippets on \SO \cite{ye2018icsd}.
In a survey with 87 Stack Overflow visitors, they reported outdated answers, wrong solutions, and buggy code. 
Their results cast light on the choice for finding programming solutions, and how often they reused a prepared solution. 
Stack Overflow had the first rank in finding solutions. 
Acar \etal conducted a comprehensive study by surveying 295 app developers, and a lab study with 54 Android developers (professionals and students) in which they were allowed to resolve coding issues with one of the following four means: any resources, Stack Overflow only, official Android documentation only, or books only \cite{acar2016you}.
Their findings suggest that developers use Stack Overflow as a major source.
Interestingly, developers who could use any resources had similar performance (functional and security correctness) to those who were assigned to use Stack Overflow only.
The lack of an official role in organizations as security champions/consultants is evident, and oftentimes this role is given to someone on the development team with limited security knowledge.
By hiring security consultants, managers can gain positive impacts from the resulting security level of products, and security testers would largely benefit from the presence of such consultants \cite{thomas2018security} \cite{poller2017can}.

\emph{\textbf{Security concern:}} %
Witschey \etal conducted a study to understand what factors affect the usage of security tools \cite{witschey2015quantifying}.
Strangely enough, being more concerned about security did not lead to greater security tool usage while having training or academic background in the security field did.
Research indicates that some organizations use external resources,\eg penetration testers, to encourage developers to pay extra attention to security in development, however, without strong support, the motives tend to lose priority compared to the important functional requirements \cite{xiao2014social}. 
Likewise, managers sometimes are obliged to make vital decisions, such as releasing the code with some known problems, due to business forces~\cite{thomas2018security}.

\emph{\textbf{Security training:}} The need for regular information security training is undeniable in companies \cite{valentine2006enhancing}.
From the training frequency viewpoint, quarterly security awareness training is recommended to renew employee knowledge concerning the latest threats and trends, and in case some difficulties exist, biannual training could be the minimum required time frame \cite{gardner2014building}.
Puhakainen \etal stressed that information security trainings and communication efforts should be continuous and integrated into the organization's usual communication efforts otherwise security policies lose their efficacy \cite{puhakainen2010improving}.
According to the SANS Institute, a security awareness program should consider 
who is going to be in the training course, which topics are suitable for the audience, and ultimately how participants engage
in order to identify how frequent security training should take place.\footnote{\url{https://www.sans.org/security-awareness-training/blog/wrong-question-how-long-should-security-awareness-training-be}}

By studying the literature we found clear evidence to corroborate the findings of this study. 
Each of the discussed studies either solely explored one factor and obtained similar results or they emphasized the importance of the studied factor to improve the state of developers in security or cryptography.
We believe that even though 81\% of the participants, as well as 75\% of their companies, are utterly concerned, \ie \emph{important} or \emph{very important}, about security, the practices of the participants and companies do not accord with their grave security concern.
However, conducting a survey has some inherent limitations.
To profoundly investigate this matter, we plan to conduct interviews with some of the participants who provided their email addresses to inspect organizational policies, project-level limitations, and objectives.. 
\section{Threats to validity}
\label{sec:threat}
Our sample of software developers using crypto APIs on \GH is limited in size. 
To increase the number of such developers, more crypto open-source projects need to be identified, and associated developers must be extracted.
In the survey, there was no participant who reported \emph{no knowledge of cryptography}, and we did not exclude any participant from our analysis.
All the participants had used crypto APIs in Java open-source projects, and it was unexpected to receive responses indicating no knowledge of cryptography.
Developers, in general, may have different viewpoints on how to evaluate their knowledge in a specific area, such as cryptography.
To lower the risk of bias assumption, we provided the participants in the survey with an extra definition of what each level of knowledge is intended to mean, and the four-level knowledge used in a previous study \cite{nadi2016jumping}.
To grasp the real knowledge of developers, we need to judge developer knowledge based on their real performance, \ie in a controlled experiment. 
We only asked the participants about their companies' practices since we did not intend to ask about the names or web addresses. 
Even though we received 97 responses from 1103 potential participants,  it may be possible that more than one participant refers to the same company.

\section{Conclusions}
\label{sec:conclusion}
We surveyed 97 developers, who used cryptography in open-source projects, and studied their security and cryptography practices.
Our analyses demonstrate that high-profile developers reported better to the developer- and company-level questions, \eg security tool usage, and background in IT security. 
It should be recalled that over 70\% of the participants and their companies are utterly concerned about security.
Nevertheless, a number of worrisome patterns, \eg lack of regular security training and security consultants and low rate of security tool usage, were observed in other participants' responses.
The results provide corroborative evidence supporting the outcome suggested by prior research.
To further understand the root causes of developer practice in this area, future studies should consider organizational policies, project-level limitations and objectives, and developer expertise in practice.

\section{Acknowledgments}
We gratefully acknowledge the financial support of the Swiss National Science Foundation for the project
``Agile Software Assistance'' (SNSF project No.\ 200020-181973, Feb.\ 1, 2019 - April 30, 2022).


\bibliographystyle{IEEEtran}
\bibliography{sample-base}

\end{document}